# Photocurrent Enhancement of Graphene Photodetectors by Photon Tunneling of Light into Surface Plasmons

Alireza Maleki[1,2], Benjamin P. Cumming[1,3], Min Gu[1,3], James E. Downes[2],

David W. Coutts[1,2], and Judith M. Dawes[1,2a)]

[1]*ARC Centre for Ultrahigh bandwidth Devices for Optical Systems (CUDOS)*

[2]*MQ Photonics Research Centre, Dept. of Physics and Astronomy, Macquarie University, Sydney 2109, Australia*

[3]*Laboratory of Artificial-Intelligence Nanophotonics, School of Science, RMIT University, Melbourne, 3001, Australia*

Abstract: We demonstrate that surface plasmon resonances excited by photon tunneling through an adjacent dielectric medium enhance photocurrent detected by a graphene photodetector. The device is created by overlaying a graphene sheet over an etched gap in a gold film deposited on glass. The detected photocurrents are compared for five different excitation wavelengths, ranging from $\lambda_0 = 570$ nm to $\lambda_0 = 730$ nm. The photocurrent excited with incident *p*-polarized light (the case for resonant surface plasmon excitation) is significantly amplified in comparison with that for s-polarized light (without surface plasmon resonances). We observe that the photocurrent is greater for shorter wavelengths (for both *s* and *p*-polarizations) due to the increased photothermal current resulting from higher damping of surface plasmons at shorter wavelength excitation. Position-dependent Raman spectroscopic analysis of the optically-excited graphene photodetector indicates the presence of charge carriers near the metallic edge. In addition, we show that the polarity of photocurrent switches across the gap as the incident light spot moves across the gap. Graphene-based photodetectors offer a simple architecture which can be fabricated on dielectric waveguides to exploit the plasmonic photocurrent enhancement of the evanescent field for detection. Applications for these devices include photo-detection, optical sensing and direct plasmonic detection.

___________________________

[a)] Author to whom correspondence should be addressed. Electronic mail: Judith.dawes@mq.edu.au.

# Introduction

Graphene, a single layer hexagonal lattice of sp$^2$ carbon molecules, is a two-dimensional (2D) material which is ideally suited for photonic and optoelectronic applications [1, 2]. In particular, graphene photodetectors have been widely investigated [3-9] because of the distinct optical and electronic properties of graphene; its ultra-wideband absorption window (single-layer graphene exhibits a constant 2.3% absorption independent of the incident light wavelength) and exceptionally fast carrier transport [10]. Graphene-based photodetectors typically use a metal-graphene design [5, 11] in which the photo-response results from both the photovoltaic and photo-thermoelectric effects, and the work function difference of metal and graphene at the contact points creates p-n junctions adjacent to the contact points by shifting the graphene Fermi level [12-19]. In the photovoltaic effect, the difference in the work functions of the metal and graphene creates built-in fields to separate the generated electron-hole pairs [12, 15, 20]. In contrast, in the thermoelectric effect, the local temperature gradients of the irradiated graphene-metal structure create thermoelectric fields due to the Seebeck effect [18, 21, 22].

However, the photoresponsivity of graphene photodetectors is usually limited. Although the 2.3% absorption of graphene is very high for just a single layer of atoms, it is small in absolute terms and this limits the overall responsivity of graphene photodetectors. In addition, the extraction of photoelectrons is normally inefficient as only a small area of the p-n junction (at the graphene-metal interface) contributes to the photocurrent generation. One approach to enhancing the photoresponsivity of graphene photodetectors is to couple light into surface plasmons with a resulting field enhancement in the graphene due to the plasmonic oscillations [3, 6].

Optical excitation of surface plasmons originates from optically-induced oscillations of free charges (conduction electrons) bound at the metal-dielectric interface. One technique to couple light to plasmonic oscillations is to diffract light by plasmonic nanostructures [23]. A variety of plasmonic nanostructures have been used to diffract and couple light into surface plasmons to improve the photocurrent generation by exploiting the significant local field enhancement near the graphene-metal contacts [6, 24-26]. However, when coupling light via plasmonic nanostructures, only a small area of the effective graphene-metal interface tends to contribute to the photocurrent generation. In addition, photodetectors based on plasmonic nanostructures often exploit resonant frequencies, and thus cannot be employed for broadband detection.

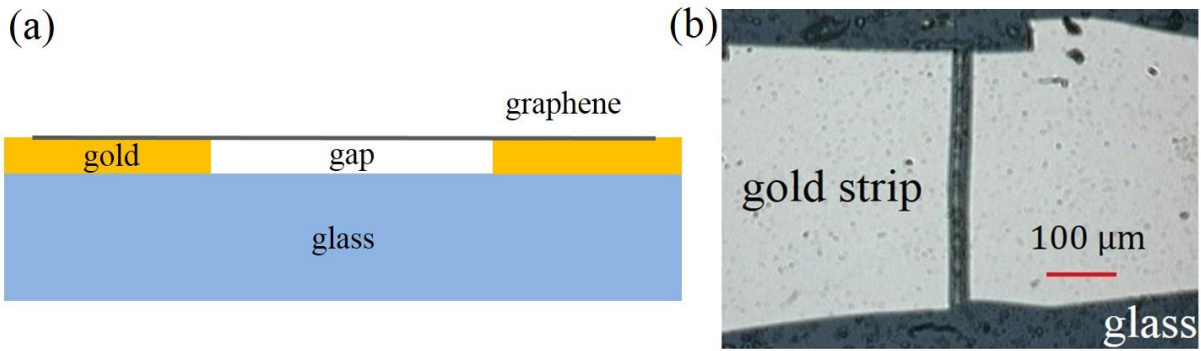

Fig. 1 (a) schematic of the graphene gap architecture, (b) image of the structure by microscope

Another technique to couple light into plasmonic oscillations is to tunnel photons into the metal-dielectric interface through an adjacent dielectric medium [23]. We show here that coupling light into surface plasmons by photon tunneling provides an effective approach to enhance photocurrent generation of a graphene photodetector. Our graphene-based photodetector consists of a 30 μm gap etched into a gold strip, with an overlaid graphene sheet linking the gold strips across the gap, see Fig. 1 (a,b). We observe photocurrent enhancement when the edges of the gold-graphene interfaces beside the gap are illuminated with polarized light ($p$-polarization) through a dielectric prism at the surface plasmon coupling angle.

Electrical detection of surface plasmons is important in the development of advanced active plasmonic devices. A variety of plasmonic detectors have been studied: surface plasmons propagating in a plasmonic device or circuit may be detected electrically, for example by inducing local changes in the resistivity of a superconducting nanowire to detect single plasmons [27], or detecting propagating surface plasmons at the Schottky contact of a semiconductor nanowire with a metal strip [28]. Coupling between a waveguide and an integrated plasmonic semiconductor-based detector allows detection of the propagating plasmons [29]. Surface plasmon leakage radiation can be similarly detected by absorption in a semiconductor substrate [30]. Our results suggest the application of graphene for electrical detection of propagating surface plasmons on metallic waveguides.

In addition, the use of a dielectric medium for enhancing photocurrent generation by graphene suggests the integration of the proposed graphene photodetector onto dielectric waveguides utilizing the evanescent field of the waveguides for electrical detection of surface plasmons which could find useful applications in sensing and photodetection. Sensing using surface plasmon waveguide sensors relies, in principle, on monitoring the change in optical parameters [31-33], however, the studied graphene photodetector can change the sensing process to an electro-optical measurement when deposited on a dielectric waveguide.

# Experimental techniques: Fabrication

The gold electrodes of the graphene gap-photodetector are made by creating a gap in a 30 nm thick gold film, which is deposited on a glass coverslip. The gold film is deposited on the 0.12 mm glass substrate (refractive index n=1.5) using an oxygen-plasma-assisted thermal deposition system to ensure good adhesion, flatness and uniformity of the pure gold film on the glass substrate [34, 35] (No additional adhesion layer is required). In order to spatially separate the excited electron- hole pairs a junction in the gold strip is required [12]. In our device, a gap is scribed in the middle of the gold strip and the graphene sheet is then positioned over the gap using Trivial Transfer Graphene™ from Advanced Chemicals Supplier. In this technique, the structure is immersed in water, and the floating PMMA-coated graphene sheet adheres to the gold, resulting in the structure depicted in Fig. 1. The sample is left for a few hours to dry and finally, the PMMA layer from the graphene is dissolved in acetone.

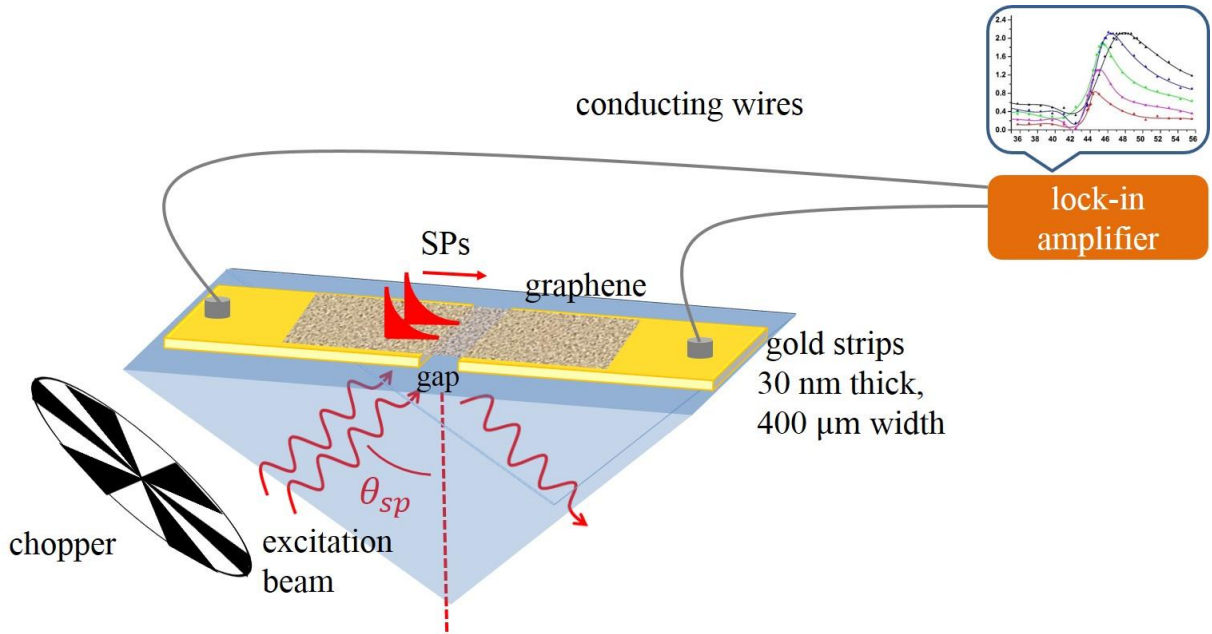

Fig.2 Schematic of the set up for the excitation of the graphene photodetector, which is contacted by index-matching oil onto a glass prism. The illumination of the structure at the surface plasmon resonance angle results in a dip in the reflected power, and also a peak in the photo-response of the detector for *p*-polarized light

# Measurement of Surface Plasmon Resonance and Photoconductance

The photodetector is placed on a glass prism (N-BK7; $n_p = 1.52$) using a refractive-index matched immersion oil, see Fig. 2. The edges of the gold-graphene interfaces at the gap are illuminated by *p*- polarized light through the prism (Kretschmann configuration for surface plasmon excitation) so that the polarized photons tunnel through the gold layer and couple into surface plasmon oscillations at the angle of incidence which corresponds to the surface plasmon resonance angle. The surface plasmon coupling angle manifests as a dip in the power of the reflected light. At this angle the projection of the wave-vector of the incident light at the metal-dielectric interface approaches the surface plasmon wave-vector according to:

$$k_{sp} = k_0 \sqrt{\varepsilon_{prism}} \sin\theta \qquad (1)$$

in which $k_{sp}$ is the surface plasmon wave vector at the gold-air interface, $k_0$ the incident light wave vector, $\varepsilon_{prism}$ the dielectric constant of the prism, and $\theta$ is the surface plasmon coupling angle [23]. The excitation light is obtained by a tunable filter (10 nm bandwidth) from a white light supercontinuum source (Fianium WhiteLase) and is tuned from 570 to 730 ± 5 nm. The excitation light is pulsed at 40 MHz repetition rate, with 10 ps pulse width, and the incident power is 1 mW. To measure the photocurrent, a chopper at frequency of ~400 Hz modulates the incident light beam and triggers a lock-in amplifier (Stanford Research System SR510). The incident light illuminates the graphene photodetector through the prism, see Fig. 2.

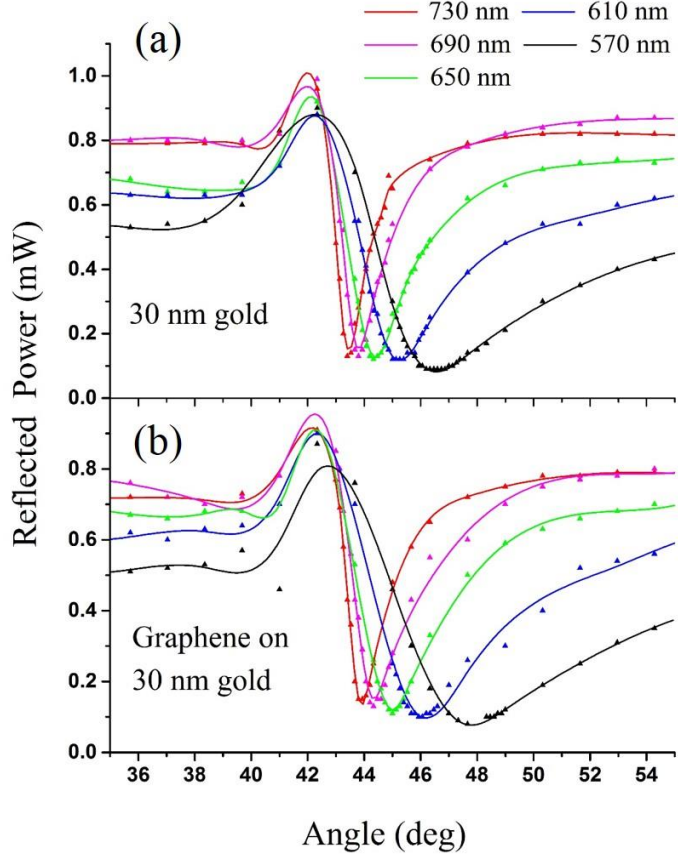

Fig. 3 reflected optical power for *p*-polarized light, plotted against the angle of incidence for (a) gold film, and (b) graphene on gold, showing the plasmon resonance angle for five different wavelengths. The data points are experimental and the lines are smooth fits.

## Results: Surface Plasmon Resonance and Photoconductance

First, the surface plasmon resonance coupling angle is measured for several excitation wavelengths for the gold layer and the gold/graphene overlaid structure, see Fig. 3. The measurements are performed at five different wavelengths, namely 730 nm, 690 nm, 650 nm, 610 nm, and 570 nm. The results in Fig. 3(a) show that with increasing wavelength of the incident polarized light, the surface plasmon resonance angle increases. We also measure the surface plasmon resonance angle for the graphene on the gold interface, shown in Fig. 3(b). The comparison between the reflection of the incident beam for the gold-air interface and the gold-graphene-air interface shows that in the latter, the addition of graphene slightly increases the surface plasmon resonance angle compared with gold alone, consistent with expectations for a slight increase in the refractive index of graphene/air compared with just air. However, this shift decreases for increasing incident light wavelength, ranging from $\Delta\theta = +1.0°$ for $\lambda_0 = 570$ to $\Delta\theta = +0.5°$ for $\lambda_0 = 730$, in agreement with the results from [36]. In addition, Fig. 3(a,b) shows that by decreasing the wavelength of incident light, the width and depth of the reflection curves increase, indicating the increase in the damping of plasmon oscillations in the metal. In fact the width of the reflection curves, is proportional to the ratio of the imaginary part of the metal dielectric constant ($\varepsilon_{im}$) to the square of its real part ($\varepsilon_{rm}$). This ratio ($\frac{\varepsilon_{im}}{(\varepsilon_{rm})^2}$) increases on moving to shorter wavelengths resulting in the increase of the width of the reflection [37-39].

The measured photocurrent against the angle of incidence for five different wavelengths shows that for incident *p*-polarized light, the generated photocurrent by graphene enhances and reaches peak values around the surface plasmon resonance angles previously measured for each of the wavelengths, see Fig. 4(a). On switching the polarization of the incident light to *s*-polarization, while maintaining the intensity of the beam at the same level as for the *p*-polarized light (1 mW), the generated photocurrent at each wavelength exhibits monotonic behavior and does not show any signal enhancement within the range of angles measured, as shown in Fig. 4(b). Both the dependence of the photocurrent enhancement on the polarization of the incident beam and also the occurrence of the enhancement around the surface plasmon resonance angle for each wavelength confirm the role of plasmon resonances in the enhancement of photocurrent generation in the graphene. The width of the photocurrent curves in Fig. 4a increases at shorter excitation wavelengths consistent with the increase in the width of the surface plasmon resonance curves in Fig. 3(a,b). This is the consequence of higher damping of plasmonic oscillations at shorter wavelengths. In addition, the measured photocurrents in Fig. 4 (a, b) also show that, on moving to shorter wavelengths, the photocurrent level increases generally for both *p* and *s*-polarizations. These are the consequence of higher damping of plasmon oscillations in gold when moving to shorter wavelengths. However, the photocurrent enhancement as the ratio of the peak of the photocurrent generated with the *p*-polarized light to that of the *s*–polarized light decreases on moving to shorter wavelengths, from 8 times for $\lambda_0 = 730$ nm to 4 times for $\lambda_0 = 570$ nm. We surmise that the different impacts of photovoltaic and photothermal effects on the enhancement can vary depending on the wavelength of the exciting light.

Higher damping of plasmonic oscillations in the gold film increases the generation of photothermal current by increasing the temperature of the gold. However, the interaction of surface plasmon oscillations with the graphene results in the generation of more photocurrent through the photovoltaic effect. Since we observe an increased photocurrent for both *p* and *s*-polarizations with decreasing wavelength of the incident light, this indicates that the photo-thermoelectric effect dominates the photovoltaic effect. In fact, structures with suspended graphene over the junction show a stronger thermoelectric effect in comparison to those deposited on polar substrates, which can be attributed to the elimination of the dominant electron cooling channel via the surface phonons of the polar substrates [21]. However, the higher enhancement of photocurrent on moving to longer excitation wavelengths is the result of the increased photovoltaic effect due to reduced surface plasmon loss in gold at longer wavelengths.

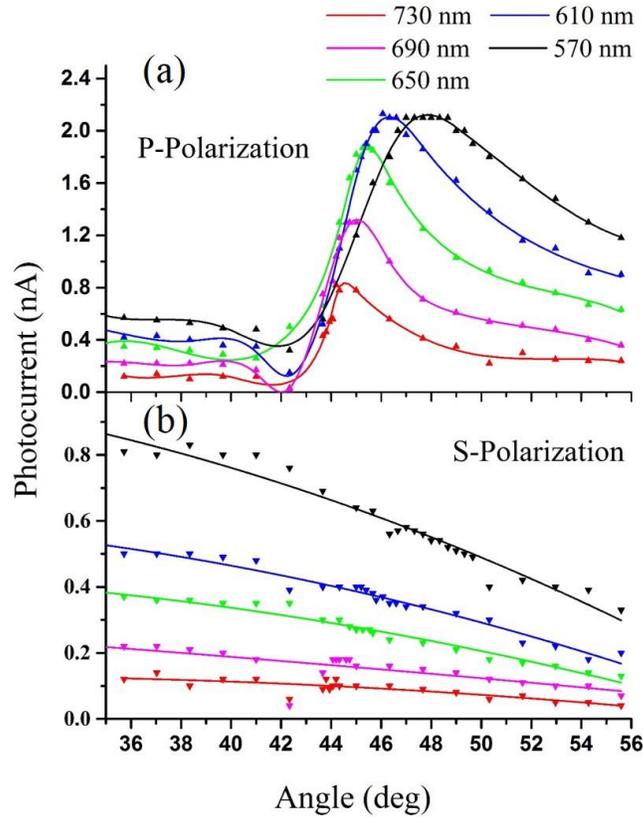

Fig. 4 The measured photocurrent of the graphene photodetector when illuminated by *p*-polarized light (a) and *s*-polarized light (b) at the wavelengths of 730, 690, 650, 610, and 570 nm. The data points are experimentally measured and the lines are smooth fits to the data points

The polarity of the photocurrent response of the graphene-based photodetector changes as the incident beam position moves across the gap, see Fig. 5(a). The incident beam spot size is large (tens of micrometers) and the angle of incidence is fixed at the relevant surface plasmon resonance angle for each wavelength when the photocurrent response is scanned. When the beam spot is moved to the first edge of the gap, the photocurrent increases from zero and reaches its maximum around the first edge of the gold-graphene interface. When the beam moves to the center of the gap the photocurrent decreases, and is zero when the center of the beam spot is positioned symmetrically at the center of the electrode gap (over the graphene). As the beam moves further towards the second edge, the photocurrent gradually increases, but with opposite polarity, and it again reaches a maximum (in absolute terms) at the gold-graphene contact edge. Moreover, the maximum photocurrent (at the contact points) exhibits a linear dependence on the incident light intensity as depicted in Fig. 5(b). These two observations, the change in the photocurrent polarity across the gap and the linear response of the structure to incident light intensity, are consistent with those reported for graphene transistors by submicron photocurrent scanning [12, 15, 17] and can be understood within the simple model of graphene band bending near the metal contacts [14, 16]. The work function difference of the metal and graphene shifts the graphene Fermi level and hence creates a potential step. Illumination of the areas close to the metal-graphene contact edge causes the photo-excited electrons to drift towards the nearby metal electrode and causes the holes to drift towards the bulk graphene, producing a photocurrent with opposite polarity at the opposite electrodes.

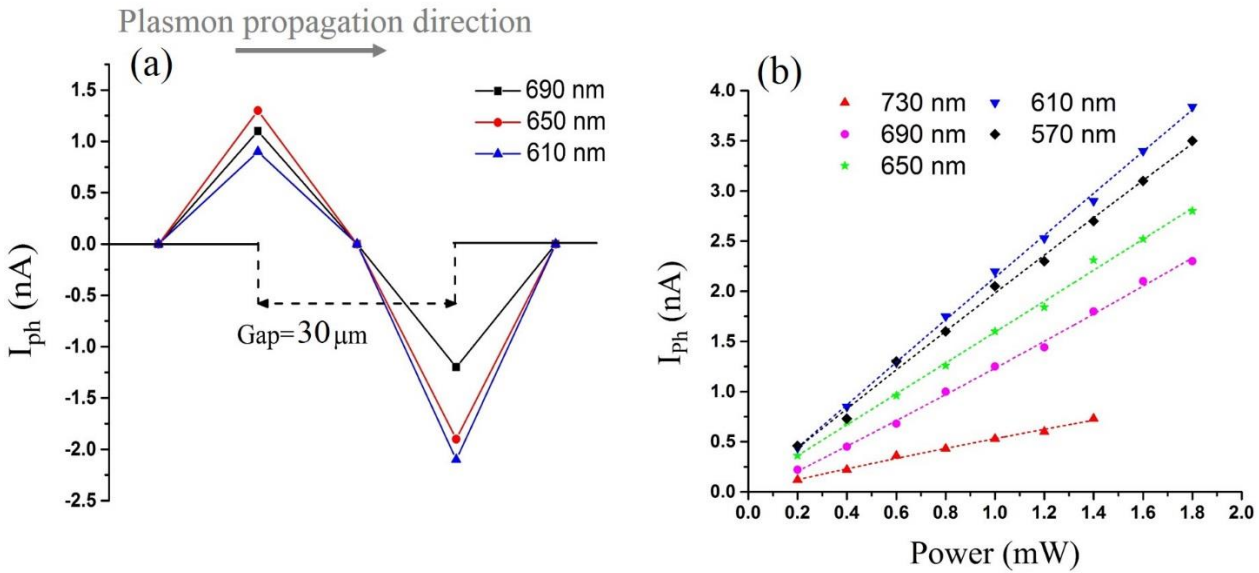

Fig. 5 (a) the change of the photocurrent polarity across the electrode gap, (b) the dependence of the photocurrent on incident light power at the surface plasmon resonance angle for each wavelength. The excitation beam is around one of the graphene-gold edges which results in maximum photocurrent enhancement. The data points are experimental results and the lines show smooth fit to the data points.

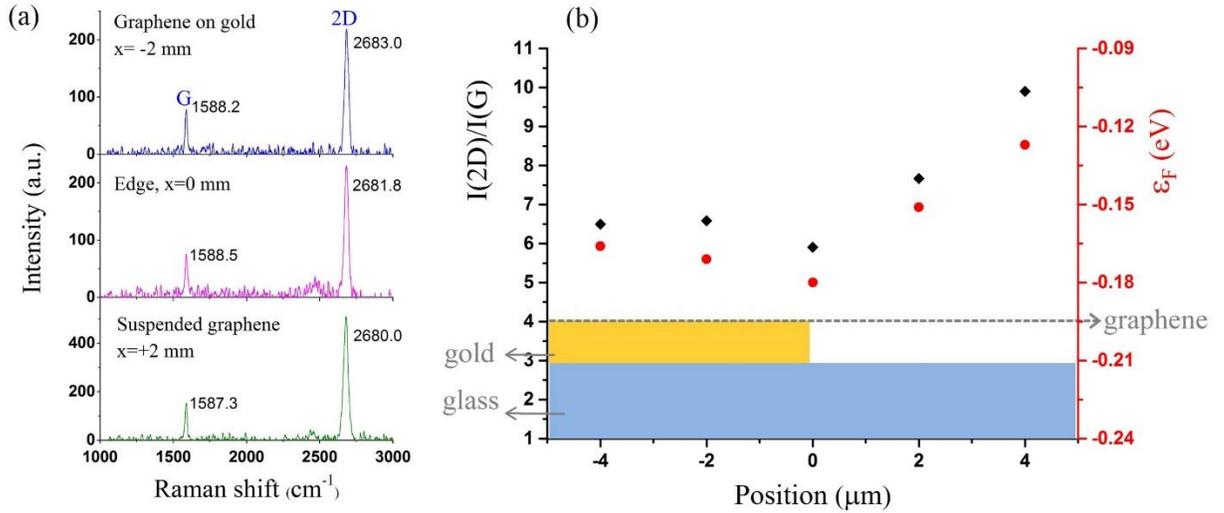

Fig. 6 (a) Raman spectra for three positions near the gold-graphene contact edge, (b) the black diamonds show the measured ratio of the intensity of the 2D peak to the G peak (I(2D)/I(G)), and red circles show the calculated Fermi level according to the wavelength shift of the G peak.

## Measurements and Analysis of Raman Spectra

Raman spectroscopy is used as a fast and non-destructive tool to identify the graphene on the structure, and to characterize the doping level and presence of defects in the graphene, in particular near the gold-graphene contact points [40-42]. The Raman spectra were obtained using a Renishaw Micro-Raman Spectrometer 2000, by illuminating different spots along the structure through a 50 × microscope objective (NA 0.75) at $\lambda_0 = 532$ nm, shown for three points in Fig. 6(a).

The main Raman spectral features of interest for our graphene structure are the G, D, and 2D peaks. The D peak is the breathing mode of the graphene carbon rings around 1360 cm$^{-1}$ and would normally only be Raman active in the presence of a defect in the graphene. Due to the absence of the D peak in the measured Raman spectra, we conclude that there are no significant defects or disorder in the graphene in our device. The G peak around 1584 cm$^{-1}$ belongs to the high-frequency E$_{2g}$ phonon at the Brillouin zone center. The position of the G peak is sensitive to the absolute value of the Fermi level ($|\varepsilon_F|$) and increases correspondingly with the increase in the Fermi level [41-44]. The shifts in the Fermi level for areas close to the gold-graphene interfaces are determined based on experimental results of monitoring dopants by electric gating of a graphene transistor [44] and depicted as red circles in Fig. 6(b). The results show that the higher absolute value of the Fermi level at the gold-graphene interface becomes a maximum at the edge of the interface (-0.18 eV) and the Fermi level decreases at the graphene suspended across the gap, indicating the potential step at the gold-graphene contact edge.

In addition to the blue shift in the position of the G peak, the shift in the position of the 2D peak can be used to distinguish between electron and hole doping in the graphene. The 2D peak ($\approx$ 2670 cm$^{-1}$), is typically the strongest Raman peak in single layer graphene, and arises as a second order Raman overtone of the D peak. Since it does not require a structural defect for its excitation, it shows a stronger dependence on Fermi level changes. The 2D peak position increases for p-doping and decreases for n-doping [45]. The Raman spectra shown in Fig 6 (a) were measured at points around the graphene-gold contact edge (-2μm, 0μm, and +2μm with the positions shown in Fig. 6(b)). There is an upshift of the 2D peak for the region around the gold graphene interface which indicates p-doping of graphene [41, 42, 44, 45].

The ratio of the intensity of the 2D and G peaks (I(2D)/I(G)) allows the change in the doping level to be identified. It has been demonstrated that by increasing the Fermi level, I(2D)/I(G) decreases [41-44]. The I(2D)/I(G) ratio for a few points around the graphene gold region is depicted as black diamonds in Fig.6(b). While the I(2D)/I(G) ratio is very similar for the graphene-gold interfaces at -4μm and -2μm from the graphene-gold edge, it decreases at the edge and increases by moving +2μm and +4μm from the edge, indicating the potential step around the edge of the graphene-gold contact and consistent with the Fermi level shifts calculated from the blue shift in the G peak position.

## Conclusions

In summary, a prism is used to enable *p*-polarized photons to tunnel into surface plasmon resonances to enhance photoexcitation in a graphene-based photodetector. Photocurrent enhancement for *p*-polarized photons at 5 different wavelengths is demonstrated as a peak at the corresponding surface plasmon resonance angles for each wavelength, while for s-polarized light at the corresponding wavelengths the photocurrents exhibit monotonic (non-resonant) behavior. In addition, the observed enhancement peaks follow the behavior of the corresponding reflectance curve for each wavelength, which governs the strong surface plasmon resonance effect on the photodetector response. Although the photocurrent enhancement decreases for smaller wavelengths, the photocurrent is in general larger for shorter wavelengths due to the stronger photothermal effect. We note that the studied graphene photodetector is not optimized and some geometrical characteristics of the structure, in particular, the gap size and the size of the graphene layer, would affect the photocurrent generation [8, 15, 17]. We believe that optimization of the graphene-based photodetector can improve the plasmonic enhancement further, through photon tunneling of light into surface plasmons. These simple devices can be used as plasmonic detectors, or can be fabricated on dielectric waveguides to exploit the waveguide evanescent field, providing a range of potential applications from photodetection to biosensing.

## Acknowledgment

We acknowledge funding and support from the ARC Centre of Excellence Program, Centre for Ultrahigh bandwidth Devices for Optical Systems (CUDOS), Swinburne University of Technology, and Macquarie University.